\documentclass[conference]{IEEEtran}
\IEEEoverridecommandlockouts

\usepackage{amsmath}
\usepackage[utf8]{inputenc} 
\usepackage{amsmath, amssymb, amsfonts} 
\usepackage{textcomp} 

\usepackage{graphicx} 
\usepackage[margin=1in]{geometry} 
\usepackage{multicol} 
\usepackage{makecell, multirow, tabularray, array} 

\usepackage{listings} 
\usepackage{xcolor} 

\usepackage{tikz} 
\usetikzlibrary{matrix, shapes, arrows, positioning, chains, calc} 
\usepackage{pgfplots} 
\pgfplotsset{compat=1.17} 

\usepackage{lipsum} 
\usepackage{cite} 
\usepackage{algorithmic} 

\definecolor{dkgreen}{rgb}{0,0.6,0}
\definecolor{gray}{rgb}{0.5,0.5,0.5}
\definecolor{mauve}{rgb}{0.58,0,0.82}
\definecolor{color1}{RGB}{175, 225, 218}
\definecolor{color2}{RGB}{232, 191, 125}
\definecolor{color3}{RGB}{168, 211, 238}
\definecolor{color4}{RGB}{97, 76, 151}

\tikzstyle{startstop} = [rectangle, text width=3cm, minimum height=1cm, text centered, draw=black, fill=red!30]
\tikzstyle{io} = [trapezium, trapezium left angle=70, trapezium right angle=110, text width=3cm, minimum height=1cm, text centered, draw=black, fill=blue!30]
\tikzstyle{process} = [rectangle, text width=3cm, minimum height=1cm, text centered, draw=black, fill=orange!30]
\tikzstyle{decision} = [diamond, text width=3cm, minimum height=1cm, text centered, draw=black, fill=green!30]
\tikzstyle{arrow} = [thick, ->, >=stealth]

\def\BibTeX{{\rm B\kern-.05em{\sc i\kern-.025em b}\kern-.08em T\kern-.1667em\lower.7ex\hbox{E}\kern-.125emX}}

\begin{document}

\newcommand{\ket}[1]{\left| #1 \right>}

\title{Novel Long Distance Free Space Quantum Secure Direct Communication for Web 3.0 Networks\thanks{}}
\author{

\IEEEauthorblockN{Yifan Zhou*}
\IEEEauthorblockA{
\textit{University of California, Los Angeles}\\
Los Angeles, United States \\
yzhou05@ucla.edu}
\\
\and
\IEEEauthorblockN{Xinlin Zhou}
\IEEEauthorblockA{
\textit{BASIS International School Guangzhou}\\
Guangzhou, China \\
xinlin.zhou13495-bigz@basischina.com}
\\
\and
\IEEEauthorblockN{Zi Yan Li}
\IEEEauthorblockA{
\textit{BASIS International School Guangzhou}\\
Guangzhou, China \\
ziyan.li11716-bigz@basischina.com}
\\
\and
\IEEEauthorblockN{Yew Kee Wong}
\IEEEauthorblockA{
\textit{Hong Kong Chu Hai College}\\
Hong Kong, China \\
ericwong@chuhai.edu.hk}
\\
\and
\IEEEauthorblockN{Yan Shing Liang}
\IEEEauthorblockA{
\textit{New York University}\\
New York, United States \\
yanshing.liang40486-bigz@basischina.com}
\\

}

\maketitle


\begin{abstract}
With the advent of Web 3.0, the swift advancement of technology confronts an imminent threat from quantum computing. Security protocols safeguarding the integrity of Web 2.0 and Web 3.0 are growing more susceptible to both quantum attacks and sophisticated classical threats. The article introduces our novel long-distance free-space quantum secure direct communication (LF QSDC) as a method to safeguard against security breaches in both quantum and classical contexts. Differing from techniques like quantum key distribution (QKD), LF QSDC surpasses constraints by facilitating encrypted data transmission sans key exchanges, thus diminishing the inherent weaknesses of key-based systems. The distinctiveness of this attribute, coupled with its quantum mechanics base, protects against quantum computer assaults and advanced non-quantum dangers, harmonizing seamlessly with the untrustworthy tenets of the Web 3.0 age. The focus of our study is the technical design and incorporation of LF QSDC into web 3.0 network infrastructures, highlighting its efficacy for extended-range communication. LF QSDC is based on the memory DL04 protocol and enhanced with our novel Quantum-Aware Low-Density Parity Check (LDPC), Pointing, Acquisition, and Tracking (PAT) technologies, and Atmospheric Quantum Correction Algorithm (AQCA). Utilizing this method not only bolsters the security of worldwide Web 3.0 networks but also guarantees their endurance in a time when quantum and sophisticated classical threats exist simultaneously. Consequently, LF QSDC stands out as a robust security solution, well-suited for Web 3.0 systems amidst the constantly evolving digital environment.
\end{abstract}



\section{Introduction}

The advent of Web 3.0 has dramatically transformed internet usage and application interaction, characterized by a decentralized, distributed, and user-centric approach. This paradigm shift empowers users with unprecedented control over their data, identity, and privacy. In the context of our globalized digital ecosystem, securing these networks is crucial, given the varied reliability and risk profiles across nodes and users. Traditional security mechanisms, including public key encryption and digital signatures, play a pivotal role in ensuring the confidentiality, integrity, and authenticity of network transactions and data. However, the emergence of quantum computing, leveraging principles of quantum mechanics, presents a formidable challenge to these cryptographic defenses, threatening to compromise data integrity by breaking current encryption methods \cite{b1,b2}.

\begin{table*}[ht]

\caption{Comparison of Types of Quantum Secure Communication Protocols}
\centering
\small
\renewcommand{\arraystretch}{1.5}
\begin{tabular}{|>{\raggedright\arraybackslash}m{2.5cm}|>{\centering\arraybackslash}m{2.5cm}|>{\centering\arraybackslash}m{2.5cm}|>{\centering\arraybackslash}m{2.5cm}|>{\centering\arraybackslash}m{2.5cm}|}
\hline
\textbf{Category} & \multicolumn{4}{c|}{\textbf{Type of Protocol}} \\
\cline{2-5} 
 & \textbf{\textit{LF QSDC}} & \textbf{\textit{DL04 Protocol}} & \textbf{\textit{Memory-Free DL04}} & \textbf{\textit{QKD}} \\
\hline
Communication Distance & \textbf{Long-distance (intercontinental)} & Moderate distance & Moderate distance & Moderate distance \\
\hline

Security Level & \textbf{High (no key exchange required)} & \textbf{High (can transmit secure messages without key exchange)} & \textbf{High (can transmit secure messages without key exchange)} & \textbf{High (key exchange fundamental) }\\
\hline
Implementation Complexity & Moderate (enhanced by PAT technologies) & High (requires quantum memory) & Moderate (no quantum memory required) & \textbf{Low (proven by ample experimentation)} \\
\hline
Suitability for Globalized Web 3.0 & \textbf{Highly suitable} & Moderately suitable & Moderately suitable & Less suitable for global scale \\
\hline
Atmospheric Disturbances Resistance & \textbf{Strong (mitigated by adaptive optics)} & Moderate (susceptible to some atmospheric effects) & Moderate (susceptible to some atmospheric effects) & Weak (highly susceptible to atmospheric effects) \\
\hline
\end{tabular}
\label{tab1}
\end{table*}

The advent of the Web 3.0 period has led to major transformations in our engagement with the Internet and its various applications. The essence of Web 3.0 lies in its decentralized, distributed, and user-centric structure, empowering users to manage their data, identity, and privacy. This scenario also paves the way for diverse economic interactions, encompassing direct peer transactions, intelligent transactions, and digital assets.

Nevertheless, the practical application of quantum communication, particularly in long-distance free-space contexts, faces significant hurdles that must be overcome to fully realize the globalization of Web 3.0 networks. Current quantum communication techniques such as QKD are unable to achieve practical secure satellite-based long-distance direct data transmission yet.

In the evolution of satellite-based QKD, significant strides have been made since the launch of the Micius satellite in 2017, demonstrating secure quantum key transmission over intercontinental distances \cite{b3,b5,b6}. These advancements include ground-to-satellite quantum teleportation \cite{b3}, air-to-ground quantum communication \cite{b4}, and satellite-based entanglement distribution \cite{b5}. However, challenges such as substantial signal loss, atmospheric interference, and the need for precise alignment and tracking remain \cite{b4,b5,b6}.

Zhou et al. addressed some of these issues with device-independent QSDC protocols \cite{b7}. Zhang et al. demonstrated QSDC over 100 km, highlighting its scalability \cite{b8}. Liu et al. discussed the infrastructure needed for global QSDC networks, including numerous satellites and ground stations \cite{b9}. Zhou et al. emphasized the importance of integrating quantum repeaters to extend QSDC's reach \cite{b10}. And more works by many scientists continue to advance the field of QSDC\cite{b11,b12,b13}

Despite many advancements, practical deployment of satellite-based QSDC still faces significant hurdles, requiring continued research and development to achieve scalable, secure quantum communication networks\cite{b14}.
\color{black}

\textbf{Challenges with QKD for Global Scale Communication:} QKD systems experience significant signal loss and attenuation over long distances, limiting practical distances without quantum repeaters to a few hundred kilometers. Atmospheric disturbances further exacerbate signal degradation in free-space QKD, making it highly susceptible to environmental factors. For instance, substantial signal loss and atmospheric interference hinder the efficiency of QKD in long-distance communications

Extending QKD reach beyond terrestrial limits with satellites introduces additional challenges, including precise satellite alignment and significant signal loss during transmission. Implementing a global QKD network requires substantial infrastructure, including numerous satellites and ground stations, and the integration of quantum repeaters. The high cost and complexity of this infrastructure pose barriers to widespread adoption

\textbf{Advantages of QSDC for Global Scale Communication} In contrast, QSDC protocols, particularly our solution LF QSDC, offer more practical solutions for global communication. QSDC allows direct transmission of secure messages without prior key exchange, reducing communication complexity and eliminating intermediate key management. Additionally, QSDC protocols incorporate eavesdropping detection through quantum state disturbance, ensuring real-time detection of interference. Technologies like Quantum-Aware Low-Density Parity-Check (LDPC) coding, Pointing, Acquisition, and Tracking (PAT) systems, and Atmospheric Quantum Correction Algorithms (AQCA) improve LF QSDC reliability over long distances. These technologies address key challenges in satellite-based free-space communication, ensuring precise alignment and minimizing signal loss due to atmospheric disturbances. LF QSDC’s reliance on advanced error correction and adaptive technologies reduces the need for extensive infrastructure compared to QKD, making it more scalable and cost-effective. By leveraging existing satellite and communication infrastructure with minimal modifications, LF QSDC can be integrated seamlessly into current systems.

Thus, while QKD has laid the groundwork for secure quantum communication, its limitations in signal loss, infrastructure requirements, and susceptibility to environmental factors make it less suitable for global-scale applications. QSDC protocols, particularly LF QSDC, present a more robust and practical solution for long-distance secure communication, offering enhanced security, scalability, and cost-effectiveness.

To address the challenge of a practical, secure, and long-distance quantum communication protocol, this research proposes a novel protocol, LF QSDC. LF QSDC is based on memory-free DL04 protocol \cite{b15} and incorporates our novel lossless transmission system which includes a Quantum-Aware LDPC coding scheme, PAT technologies, and AQCA.

Recent advancements in QSDC and supporting fields have shown significant theoretical and experimental progress, enhancing the feasibility of direct quantum communications\cite{b16,b17,b18,b19,b20,b21}. Noteworthy developments include the realization of QSDC over 100 km using time-bin and phase quantum states underlines the scalability of these technologies\cite{b10}.

The memory-free DL04 protocol contributes to these advancements by facilitating secure quantum communication without the need for a shared secret key and bypassing the necessity for quantum memory. This protocol ensures the integrity of quantum communication through immediate state preparation and encoding, followed by direct transmission to the receiver. This process is detailed as follows\cite{b10}:

\begin{enumerate}
    \item \textbf{Quantum State Preparation:} Bob prepares qubits in one of the four initial states: $\{|0\rangle, |1\rangle\}$ (polarization states) or $\{|+\rangle, |-\rangle\}$ (phase states), forming the basis of the communication.
    
    \item \textbf{Initial Transmission to Alice:} Bob sends these qubits to Alice via the quantum channel.
    
    \item \textbf{Eavesdropping Detection:} Alice randomly selects some received qubits for immediate measurement in basis $X$ or $Z$. The results are communicated to Bob through the classical service channel. Bob then verifies if the measured qubits match the initially prepared states. Any discrepancy indicates potential eavesdropping by Eve, causing the process to terminate if the error rate ($e_e$) exceeds a predefined threshold. Otherwise, Alice and Bob proceed to estimate the secrecy capacity ($C_s$).
    
    \item \textbf{Message Encoding:} Alice encodes the message bits ($M_k$) into codewords ($C_{nc}$) using a predetermined coding scheme.
    
    \item \textbf{Photon Modulation:} Alice modulates the remaining qubits by applying either the identity operator ($I$) or the unitary operator ($U$) based on the bit values '0' or '1' of $C_{nc}$. These modulated photons are then stored in a quantum memory.
    
    \item \textbf{Return Transmission to Bob:} The modulated photons are sent back to Bob through the same quantum channel.
    
    \item \textbf{Demodulation and Decoding:} Bob demodulates the received photons to retrieve the codewords ($C'_{nc}$), then decodes these to extract the original message ($Y_k$).
\end{enumerate}

\textbf{Key Features and Advantages}

\begin{itemize}
    \item \textbf{Eavesdropping Detection:} The immediate measurement of randomly selected qubits allows Alice and Bob to detect any interference by Eve, leveraging the fundamental principles of quantum mechanics where any measurement alters the quantum state.
    
    \item \textbf{Quantum Memory Utilization:} Unlike protocols that operate memory-free, DL04 requires quantum memory, ensuring qubits are stored securely between initial transmission and modulation. This aids in maintaining the integrity and sequence of transmitted qubits.
    
    \item \textbf{Secrecy Capacity Estimation:} The protocol allows for precise estimation of the secrecy capacity ($C_s$), which is crucial for determining the security of the communication channel.
    
    \item \textbf{Two-Way Transmission:} The bidirectional flow of qubits between Alice and Bob enhances the robustness of the protocol by providing an additional layer for detecting and mitigating eavesdropping attempts.
\end{itemize}

\color{black}
The memory-free DL04 protocol that FL QSDC employs reduces the complexity of quantum state storage and management, making the system more practical and robust for real-world applications \cite{b15} as shown in Table I. Our novel 1uantum-aware LDPC coding scheme, PAT technologies, and AQCA mitigate issues related to atmospheric turbulence and alignment errors, which are prominent in satellite-based free-space quantum communication. Our innovative features are based on the developments of these works \cite{b22,b23,b24,b25,b26,b27}.

This paper delves into a strategy for incorporating LF QSDC into Web 3.0 frameworks to combat quantum dangers, in harmony with the decentralized nature of Web 3.0. Integrating LF QSDC technically with Web 3.0 demands an innovative method. It's necessary to modify Web 3.0 protocols to integrate LF QSDC's direct transmission features. This demands progress in quantum communication technologies and the evolution of Web 3.0 architecture to facilitate such integration.

The main contributions of this article are summarized as follows:
\begin{enumerate}
    \item Introduction of a novel LF QSDC system designed for Web 3.0 networks.
    \item Introduction of a detailed and practical road map to the implementation of LF QSDC into global communication networks.
    \item Development and optimization of a quantum-aware LDPC and PAT technology to enhance quantum communication reliability and efficiency.
    \item Proposal of a novel AQCA aimed at mitigating atmospheric disturbances and improving security over long distances satellite communication.
\end{enumerate}
 We will first give an overview of our proposed LF QSDC system. Then, we will detail our innovative designs of quantum-aware LDPC, PAT technologies, and atmospheric quantum correction algorithms. Finally, we will present our implementation plan for the LF QSDC and discuss the implications of our findings for global communication networks.


\section{Long-Distance Free-Space QSDC Overview}

This section introduces long-distance free-space quantum secure direct communication (QSDC). Our investigation delves into how advanced technologies can surmount existing constraints, rendering QSDC viable for widespread application both in the open air and across space through satellite communication. The process initiates by setting up the quantum state, marking the first phase in encoding quantum information for transmission. Following this, the process involves applying LDPC coding to enhance error correction capabilities, crucial for maintaining the integrity of quantum data over long distances. PAT technologies are integrated to ensure precise alignment and stabilization of the quantum signal. An atmospheric correction step is then applied to mitigate effects like scattering and absorption that can degrade the quantum signal as it traverses the atmosphere. The corrected signal undergoes transmission, sending it across free space, potentially covering vast distances including satellite-to-ground communication paths. Upon reaching the destination, the quantum signal is measured in a critical phase where the encoded quantum information is detected and interpreted. Finally, the process concludes with information decoding, where the quantum data is translated back into classical information for use.

Figure 1 shows the overview of LF QSDC: 
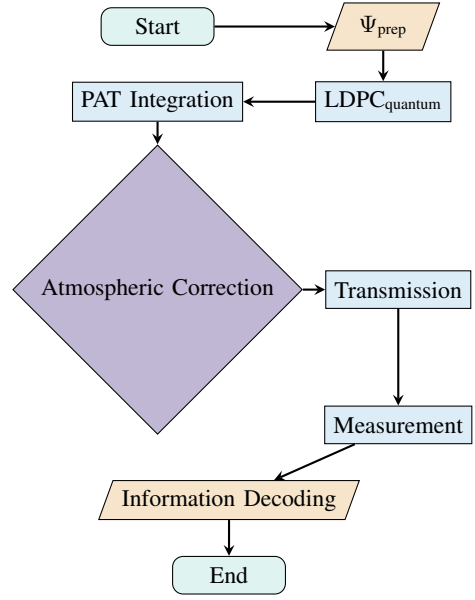
\begin{figure} [ht]
    \centering
        \begin{tikzpicture}[node distance=1cm, font = \small]
        
        \tikzstyle{startstop} = [rectangle, rounded corners, minimum width=1.5cm, minimum height=0.5cm, text centered, draw=black, fill=color1!40, font=\small]
        \tikzstyle{io} = [trapezium, trapezium left angle=70, trapezium right angle=110, minimum width=1.5cm, minimum height=0.5cm, text centered, draw=black, fill=color2!40, font=\small]
        \tikzstyle{process} = [rectangle, minimum width=1.5cm, minimum height=0.5cm, text centered, draw=black, fill=color3!40, font=\small]
        \tikzstyle{decision} = [diamond, minimum width=1.5cm, minimum height=0.5cm, text centered, draw=black, fill=color4!40, font=\small]
        \tikzstyle{arrow} = [thick,->,>=stealth]
        
        \node (start) [startstop] {Start};
        \node (in1) [io, right of=start,xshift=2cm] {$\Psi_{\text{prep}}$};
        \node (pro1) [process, below of=in1] {$\text{LDPC}_{\text{quantum}}$};
        \node (pro2) [process, left of=pro1,xshift=-2cm] {PAT Integration};
        \node (dec1) [decision, below of=pro2, yshift=-1.5cm] {Atmospheric Correction};
        \node (pro3) [process, right of=dec1, xshift=2.2cm] {Transmission};
        \node (pro4) [process, below of=pro3, yshift=-0.8cm] {Measurement};
        \node (out1) [io, below of=pro4, xshift=-2.25cm] {Information Decoding};
        \node (stop) [startstop, below of=out1] {End};
        
        \draw [arrow] (start) -- (in1);
        \draw [arrow] (in1) -- (pro1);
        \draw [arrow] (pro1) -- (pro2);
        \draw [arrow] (pro2) -- (dec1);
        \draw [arrow] (dec1) -- (pro3);
        \draw [arrow] (pro3) -- (pro4);
        \draw [arrow] (pro4) -- (out1);
        \draw [arrow] (out1) -- (stop);
        \end{tikzpicture}
    \caption{Flowchart showing the overview of FL QSDC}
    \label{fig: Overview of FL QSDC}
\end{figure}

\subsection{Quantum State Preparation and Encoding}

The LF QSDC protocol starts with Alice preparing and encoding quantum states in a highly structured manner. She uses two parameters, \(a_0\) and \(a_1\), to determine the quantum state based on the required information:

\subsubsection{Basis Determination (\(a_0\)):}
  \begin{itemize}
    \item If \(a_0 = 0\), the basis is computational, with states \(\{|0\rangle, |1\rangle\}\).
    \item If \(a_0 = 1\), the basis is superpositional, where \(a_1\) further dictates the specific state:
      \begin{itemize}
        \item \(a_1 = 0\): State becomes \(\frac{|0\rangle + |1\rangle}{\sqrt{2}}\)
        \item \(a_1 = 1\): State becomes \(\frac{|0\rangle - |1\rangle}{\sqrt{2}}\)
      \end{itemize}
  \end{itemize}

Alice also selects a third parameter, \(a_2\), which applies a phase shift to the encoded states:

\begin{itemize}
  \item If \(a_2 = 0\), the phase states remain as initially encoded.
  \item If \(a_2 = 1\), the phase states are swapped:
    \begin{itemize}
      \item \(|0\rangle + |1\rangle\) becomes \(|0\rangle - |1\rangle\) and vice versa.
    \end{itemize}
\end{itemize}

These encoded quantum states are then transmitted to Bob, who has a crucial role in their measurement.

\subsection{Measurement Protocols and Two-way Communication}

Upon receiving the quantum states, Bob's task is to measure them accurately, which requires selecting the appropriate basis for measurement based on preliminary information shared by Alice. Bob randomly selects a measurement basis, theta (\(\theta\)), which dictates his measurement strategy:

\begin{itemize}
  \item If \(\theta = 0\), he measures in the basis:
    \begin{equation}
      \{|0\rangle, |1\rangle\}
    \end{equation}

  \item If \(\theta = 1\), he measures in the basis:
    \begin{equation}
      \left\{\frac{|0\rangle + |1\rangle}{\sqrt{2}}, \frac{|0\rangle - |1\rangle}{\sqrt{2}}\right\}
    \end{equation}
\end{itemize}

After measurement, Bob announces his choice of \(\theta\) and the results of his measurements. Alice and Bob then jointly analyze the outcomes to check if \(\theta\) aligns with \(a_2\) (\(a_2 \oplus \theta = 0\) or \(1\)), ensuring the correct basis was used and the transmission was secure. This process embodies the two-way communication required by the DL04 protocol, where feedback from Bob influences potential retransmissions by Alice or adjustments in the encoding strategy to correct any discrepancies or enhance security.

\subsection{Integration of Advanced Quantum Technologies}

The effective transmission of quantum states over long distances is fraught with potential disturbances such as atmospheric turbulence, scattering, and absorption. To mitigate these effects and enhance the fidelity of the quantum channel, the following technologies are integrated into our LF QSDC protocol:
\begin{enumerate}
  \item \textbf{Quantum-Aware LDPC Coding:} Specifically designed for quantum information, these codes correct errors that occur during the quantum state transmission, thus ensuring that the integrity and secrecy of the data are maintained even over long distances.
  \item \textbf{Pointing, Acquisition, and Tracking (PAT) Systems:} These technologies are critical for maintaining the alignment of the quantum communication link, especially in dynamic environments such as satellite communications, where precision pointing is crucial for successful data transmission.
  \item \textbf{Atmospheric Quantum Correction Algorithms:} These algorithms are designed to compensate for the quantum signal degradation caused by the atmosphere. By correcting errors induced by atmospheric turbulence, these algorithms significantly improve the reliability and stability of the quantum channel.
\end{enumerate}

\subsection{Security Checking Process}

To ensure the security of the QSDC protocol, a detailed security checking process is implemented. After the initial transmission and measurement phases, a verification step is conducted where Alice and Bob compare a subset of their data to check for any discrepancies. This process includes the following steps:
\begin{enumerate}
  \item \textbf{Error Rate Estimation:} Alice and Bob share a portion of their measurement results to estimate the quantum bit error rate (QBER). If the QBER exceeds a predefined threshold, the communication is deemed insecure, and the process is aborted.
  \item \textbf{Entanglement Verification:} By verifying the entanglement of the transmitted quantum states, Alice and Bob can ensure that no eavesdropping has occurred. This involves comparing the measurement results to check for correlations consistent with the expected entangled states.
  \item \textbf{Classical Post-Processing:} Any discrepancies identified during the verification steps are corrected through classical post-processing techniques, such as error correction and privacy amplification. This ensures that the final shared data is secure and free from eavesdropping.
\end{enumerate}

By incorporating these security measures, the LF QSDC protocol ensures the secure transmission of quantum information over long distances, making it suitable for practical applications in free-space and satellite-based communication systems.

\color{black}

\section{Lossless and Secure Long-Distance Free-Space Transmission Techniques}

This section delves into our innovative techniques for the secure and loss-free transmission of data over extensive distances via air or space.

\subsection{Quantum-Aware LDPC Coding}

\subsubsection {LDPC Code Parameter Optimization}

\textbf{Degree Distribution Optimization:}
\begin{itemize}
    \item \textit{Extended Definition and Impact:} The degree distribution of an LDPC code, represented by the polynomials $\lambda(x)$ for variable nodes and $\rho(x)$ for check nodes, fundamentally determines the code's performance in terms of error correction efficiency and rate. These polynomials dictate how bits are interconnected within the LDPC graph.
    \item \textit{Technical Enhancement:} A comprehensive optimization function that not only maximizes mutual information I(X; Y) and minimizes QBER but also considers the decoding threshold and the code rate. The optimization can integrate more complex constraints related to the noise model of the quantum channel.
    \item \textit{New Mathematical Formulation:} \newline
        Optimization Goal:
        \begin{equation}
            \max \bigl( I(X; Y) - \lambda(\text{QBER}) - \delta(\text{Code Rate}) \bigr)
        \end{equation}
        Subject to: 
        \begin{equation}
            \lambda(x) = \sum_{i=2}^{d_v} \lambda_i x^{i-1}, \quad \rho(x) = \sum_{i=2}^{d_c} \rho_i x^{i-1}
        \end{equation}
        Here, $\delta$ represents the additional constraint related to the code rate.
\end{itemize}
    
\textbf{Channel Adaptation:}
\begin{itemize}
    \item \textit{Machine Learning Prediction and Iterative Update Algorithm:} a machine learning and an iterative update algorithm designed to predict upcoming alterations in the quantum channel and modify the LDPC code parameters, guided by immediate feedback from the quantum communication system.
        \begin{enumerate}
            \item Initialization: Set initial LDPC code parameters based on average channel conditions.
            \item Real-time Feedback Loop:
                \begin{itemize}
                \renewcommand{\labelitemii}{*}
                    \item Collect real-time CQ data.
                    \item Adjust LDPC parameters for immediate channel conditions.
                \end{itemize}
            \item Predictive Adjustment:
                \begin{itemize}
                \renewcommand{\labelitemii}{*}
                    \item Use the ML model to predict short-term future CQ.
                    \item Preemptively adjust LDPC parameters based on predictions.
                \end{itemize}
            \item Iterative Update:
                \begin{itemize}
                \renewcommand{\labelitemii}{*}
                    \item Continuously repeat steps 2 and 3.
                    \item Employ a decay factor to balance between recent adjustments and new predictions.
                \end{itemize}
        \end{enumerate}
    \item \textit{Pseudocode:}
        \begin{lstlisting}
# LSTM Model for Channel Prediction
def build_lstm_model(input_shape):
model = Sequential()
model.add(LSTM(50, return_sequences=True, input_shape=input_shape))
model.add(LSTM(50))
model.add(Dense(1))
model.compile(optimizer='adam', loss='mse')
return model

# Update LDPC Parameters
def update_ldpc_parameters(current_params, current_cq, predicted_cq):
# Algorithm to update LDPC parameters based on current and predicted CQ
# Example: Adjust code rate based on SNR
snr_threshold = 10
if current_cq['SNR'] < snr_threshold or predicted_cq['SNR'] < snr_threshold:
new_params = adjust_code_rate(current_params, decrease=True)
else:
new_params = adjust_code_rate(current_params, decrease=False)
return new_params

# Main Loop for Iterative Update
def iterative_update_loop(initial_params, model, channel_data):
params = initial_params
for t in range(len(channel_data)-1):
current_cq = get_current_cq(channel_data, t)
predicted_cq = model.predict(channel_data[t:t+1])
params = update_ldpc_parameters(params, current_cq, predicted_cq)
# Update the LDPC codes with new params
return params
        \end{lstlisting}
\end{itemize}

\subsubsection {Adaptive Decoding Algorithm with Graph Neural Network Enhanced Belief Propagation and Adaptive Iteration}
We propose a GNN-enhanced adaptive decoding algorithm for quantum communication, improving LDPC code decoding by adapting to channel changes and enhancing security with decoy states for eavesdropping detection and privacy amplification methods to secure fi- nal keys, significantly increasing transmission reliability and protection.

\textbf{Standard BP Equation:}
\begin{equation}
    m^{(t+1)}_{i \rightarrow j} = 2 \tanh^{-1}\left(\prod_{k \in N(i)\backslash j} \tanh\left(\frac{m^{(t)}_{k \rightarrow i}}{2}\right)\right)
\end{equation}

\textbf{Modification with GNN:} 
\\
Adjust messages \( m^{(t)}_{i \rightarrow j} \) using a GNN. The GNN predicts the likelihood of error in each node's message, influencing the BP update rule.

\textbf{GNN Model Design:}
\begin{itemize}
    \item Input: Messages from each node, current iteration number, and additional features like channel conditions.
    \item Architecture: Utilize a GNN architecture capable of handling graph-structured data. Layers can include graph convolutional networks (GCN) or Graph Attention Networks (GAT).
    \item Output: Adjusted messages and probability of error for each node.
\end{itemize}

\textbf{Adaptive Iteration:}
\begin{itemize}
    \item Dynamically Adjust Iterations: Based on channel conditions and convergence rate, the number of iterations, \( T \), is adapted.
    \item Stopping Criterion: Utilize error patterns and rate of convergence to determine when to stop the iterations.
\end{itemize}

\textbf{Mathematical Representation:}
\begin{itemize}
    \item Let \( m^{(t)}_{i \rightarrow j} \) be the message from node \( i \) to node \( j \) at iteration \( t \).
    \item GNN output influences the update rule: 
    \begin{equation}
        m^{(t+1)}_{i \rightarrow j} = GNN(m^{(t)}_{i \rightarrow j}, \text{features})
    \end{equation}

    \item Adaptive iteration count \( T \) based on GNN feedback and channel conditions.
\end{itemize}

\textbf{Pseudocode:}
\begin{lstlisting}
class GNNModel(torch.nn.Module):
    #Graph Neural Network model using GCN layers.
    def __init__(self, input_dim, hidden_dim, output_dim):
        super().__init__()
        self.conv1 = GCNConv(input_dim, hidden_dim)  # First GCN layer
        self.conv2 = GCNConv(hidden_dim, output_dim)  # Second GCN layer

    def forward(self, x, edge_index):
        x = F.relu(self.conv1(x, edge_index))  # Activation function
        return self.conv2(x, edge_index)  # Output layer
        
def enhanced_bp(messages, gnn_model, edge_index, num_iterations):
    #Enhanced BP decoding with GNN model.
    for _ in range(num_iterations):
        messages_tensor = torch.tensor(messages, dtype=torch.float)  # Messages to tensor
        messages = gnn_model(messages_tensor, edge_index).detach().numpy()  # GNN model output
    return messages

def adaptive_decoding_algorithm(messages, channel_conditions, edge_index):
    #Adaptive decoding based on channel conditions.
    input_dim, hidden_dim, output_dim = messages.shape[1], 64, messages.shape[1]
    gnn_model = GNNModel(input_dim, hidden_dim, output_dim)
    T = 10 if channel_conditions['SNR'] > 10 else 20  # Iterations based on SNR
    adjusted_messages = enhanced_bp(messages, gnn_model, edge_index, T)  # Decode with GNN
    return decode(adjusted_messages)  # Final decoding
\end{lstlisting}

\subsubsection {Security Enhancement Techniques}

\textbf{Decoy State Method:}
\begin{itemize}
    \item Employ decoy states with varying intensities to estimate channel parameters.
    \item Use statistical methods to analyze the difference in detection rates between signal and decoy states to estimate \( P_{loss} \) and detect eavesdropping.
    \item The estimation can be formulated as solving a set of linear inequalities derived from decoy state intensities and detection rates.
\end{itemize}

\textbf{Privacy Amplification:}
\begin{itemize}
   \item After decoding, apply a hash function \( H \) to the key to reduce its length and eliminate partial information\cite{b28,b29}.
   \item The process can be represented as:
        \begin{equation}
            K_{final} = H(K_{decoded})
        \end{equation}
    \item The choice of \( H \) and the length of \( K_{final} \) are critical and depend on the estimated QBER and eavesdropping probabilities.
\end{itemize}

\subsection{PAT Technologies}

\subsubsection{Design}

The PAT system is designed to automatically adjust the direction and position of quantum signal transmission and reception equipment, ensuring optimal alignment between communicating parties. This is crucial for minimizing signal loss and maintaining high fidelity of the quantum state during transmission.

The PAT system consists of three main components: a pointing device, a tracking device, and a control device. The pointing device is responsible for directing the quantum signal beam toward the intended receiver, using a combination of mechanical and optical elements, such as mirrors, lenses, and motors. The tracking device is responsible for detecting the incoming quantum signal beam from the transmitter, using a photodetector or a camera. The control device is responsible for coordinating the pointing and tracking devices, using feedback loops and algorithms, to achieve the desired alignment and stability.

The PAT system operates in two modes: a coarse mode and a fine mode. The coarse mode is used to establish the initial alignment between the transmitter and receiver, using a wide-angle beam and a low-resolution detector. The fine mode is used to refine the alignment and maintain stability, using a narrow-angle beam and a high-resolution detector.

\subsubsection{Performance Evaluation and Optimization Algorithms}

The performance of the PAT system can be quantified by several metrics, such as alignment error, signal acquisition probability, and pointing stability. These metrics can be formulated by mathematical equations, as follows:

\textbf{Alignment Error Correction:}
    
        The alignment error ($e_a$) is modeled as a function of the angular misalignment ($\theta_m$) and the distance ($d$) between the transmitter and receiver:
        \begin{equation}
            e_a = f(\theta_m, d)
        \end{equation}
        where $f(\cdot)$ represents the functional relationship, which is determined empirically or through simulation. The alignment error measures the deviation of the quantum signal beam from the ideal optical axis, which can result in signal loss or state degradation. The PAT system aims to minimize the alignment error by adjusting the pointing and tracking devices accordingly.

        The alignment error correction algorithm ($AEC$) can be expressed as:
        \begin{equation}
            AEC = \arg\min_{\theta_p, \theta_t} e_a(\theta_p - \theta_t, d)
        \end{equation}
        where $\theta_p$ and $\theta_t$ are the pointing and tracking angles, respectively. The AEC algorithm finds the optimal pointing and tracking angles that minimize the alignment error, using methods such as gradient descent or Newton's method.

\textbf{Signal Acquisition:}

        The probability of successful signal acquisition ($P_a$) depends on the signal-to-noise ratio (SNR) and the alignment precision ($\sigma$):
        \begin{equation}
            P_a = g(SNR, \sigma)
        \end{equation}
        with $g(\cdot)$ encapsulating the acquisition algorithm's efficiency under varying SNR conditions and alignment precisions. The signal acquisition probability measures the likelihood of establishing a quantum link between the transmitter and receiver, which can be affected by noise sources, such as background light, thermal noise, and dark counts. The PAT system aims to maximize the signal acquisition probability by optimizing the SNR and the alignment precision.
        
        The signal acquisition algorithm ($SAC$) can be expressed as:
        \begin{equation}
            SAC = \arg\max_{SNR, \sigma} P_a(SNR, \sigma)
        \end{equation}
        where $SNR$ and $\sigma$ are the signal-to-noise ratio and the alignment precision, respectively. The SAC algorithm finds the optimal SNR and alignment precision that maximize the signal acquisition probability, using methods such as thresholding or Bayesian inference.

 \textbf{Pointing Stability:}

        The pointing stability requirement ($S_p$) to maintain the quantum link can be expressed as:
        \begin{equation}
            S_p < \frac{\lambda}{D_{eff}}
        \end{equation}
        where $\lambda$ is the wavelength of the quantum signal, and $D_{eff}$ is the effective aperture diameter of the transmitter/receiver system. The pointing stability requirement measures the maximum allowable angular deviation of the quantum signal beam from the optical axis, which external factors, such as wind, vibration, and turbulence can cause. The PAT system aims to satisfy the pointing stability requirement by stabilizing the pointing and tracking devices against these factors.
        
        The pointing stability algorithm ($PSA$) can be expressed as:
        \begin{equation}
            PSA = \arg\min_{\Delta\theta_p, \Delta\theta_t} S_p(\Delta\theta_p, \Delta\theta_t)
        \end{equation}
        where $\Delta\theta_p$ and $\Delta\theta_t$ are the pointing and tracking angular deviations, respectively. The PSA algorithm finds the optimal pointing and tracking angular deviations that minimize the pointing stability requirement, using methods such as PID control or Kalman filter.

\subsection{Atmospheric Quantum Correction Algorithm}

\subsubsection{Design}

AQCA is integrated within the LF QSDC framework to address challenges such as atmospheric turbulence, absorption, and scattering. These phenomena can degrade the quantum state of photons, leading to increased quantum bit error rates (QBER) and reduced communication security and reliability.

The AQCA consists of four main modules: a disturbance modeler, a quantum error corrector, an adaptive optics system, and a signal enhancer and recoverer. The disturbance modeler is responsible for estimating the atmospheric parameters and their impact on the quantum signal. The quantum error corrector is responsible for applying QEC techniques to the quantum signal. The adaptive optics system is responsible for adjusting the quantum signal’s path in real-time. The signal enhancer and recoverer are responsible for processing the quantum signal to improve the SNR and recover the original quantum state.

\subsubsection{Mathematical Formulation}

The AQCA leverages advanced mathematical models and algorithms to correct for atmospheric distortions. Key components of the algorithm include:

\textbf{Modeling Atmospheric Disturbances:}
        
        The Kolmogorov theory models the atmospheric turbulence. The strength of the turbulence is characterized by the Fried parameter ($r_0$), which represents the coherence length of the wavefront. The effect of the turbulence on the quantum signal is quantified by the scintillation index ($\sigma_I^2$), which measures the intensity fluctuations of the signal. The scintillation index can be approximated by the Rytov approximation, which is valid for weak to moderate turbulence regimes. The Rytov approximation is given by:
        \begin{equation}
            \sigma_I^2 \approx 1.23 C_n^2 k^{7/6} L^{11/6}
        \end{equation}

        The atmospheric absorption is modeled by the Beer-Lambert law. The effect of the absorption on the quantum signal is quantified by the transmittance ($T$), which measures the fraction of the signal that reaches the receiver. The transmittance is given by:
        \begin{equation}
            T = e^{-\alpha L}
        \end{equation}

        The Mie theory models atmospheric scattering. The effect of the scattering on the quantum signal is quantified by the scattering cross section ($\sigma_s$), which measures the probability of the signal being scattered by a particle. The scattering cross section is given by:
        \begin{equation}
            \sigma_s = \frac{2\pi^5 d^6}{3\lambda^4} \left(\frac{n^2-1}{n^2+2}\right)^2 Q_{ext}
        \end{equation}

\textbf{Quantum Error Correction (QEC):}

        The AQCA employs QEC techniques tailored to atmospheric conditions. These techniques are designed to identify and correct errors induced by the atmosphere, enhancing the resilience of quantum communication.
        
        The AQCA selects the appropriate type of quantum code based on the quantum signal's modulation scheme. For phase-modulated signals, such as coherent states or squeezed states, the AQCA uses CV codes, such as Gaussian codes or non-Gaussian codes. For polarization-modulated signals, such as single photons or entangled photons, the AQCA uses DV codes, such as stabilizer codes or non-stabilizer codes.

        The QEC process consists of three steps: encoding, syndrome measurement, and decoding. Encoding is the process of applying a quantum code to the quantum signal before transmission. Syndrome measurement is the process of measuring the quantum signal after transmission to detect errors. Decoding is the process of applying a quantum code to the quantum signal after syndrome measurement to correct the errors.

\textbf{Adaptive Optics System:}

        An adaptive optics system is integrated to dynamically adjust the quantum signal's path in real time, countering the effects of atmospheric turbulence. The system uses feedback from the quantum signal itself to optimize the transmission path.
        
        The adaptive optics system consists of three main components: a \textit{wavefront sensor}, a \textit{deformable mirror}, and a \textit{control unit}. The wavefront sensor is responsible for measuring the phase distortions of the quantum signal caused by the turbulence. The deformable mirror is responsible for compensating the phase distortions by applying a conjugate phase profile to the quantum signal. The control unit is responsible for coordinating the wavefront sensor and the deformable mirror, using feedback loops and algorithms, to achieve the optimal wavefront correction.
        
        The adaptive optics system operates in two modes: a\textit{ closed-loop mode} and an \textit{open-loop mode}. The closed-loop mode is used when the quantum signal is strong enough to provide sufficient feedback for the wavefront sensor. The open-loop mode is used when the quantum signal is too weak to provide sufficient feedback for the wavefront sensor. In this case, the system uses a reference beam, such as a laser beam, to provide the feedback for the wavefront sensor, and applies the same correction to the quantum signal.
        
\subsection{Integration with LF QSDC Protocol}
The integration of PAT technologies with the QSDC protocol involves the synchronization of quantum state preparation, signal transmission, and reception processes with the dynamic adjustments made by the PAT system. 

The integration process can be summarized by the following steps: 

\begin{enumerate}
    \item The transmitter and the receiver measure the atmospheric parameters, such as the Fried parameter, the absorption coefficient, and the scattering cross section, using the disturbance modeler module of the AQCA.
    \item The transmitter and the receiver select the appropriate quantum code, such as a CV code or a DV code, based on the quantum signal’s modulation scheme and the atmospheric parameters, using the quantum error corrector module of the AQCA.
    \item The transmitter encodes the quantum state into the quantum signal, using a quantum source and a quantum modulator, and applies the quantum code to the quantum signal, using the encoding step of the QEC process.
    \item The transmitter directs the quantum signal toward the receiver, using the pointing device of the PAT system, and adjusts the quantum signal’s path in real-time, using the adaptive optics system module of the AQCA.
    \item The receiver detects the incoming quantum signal, using the tracking device of the PAT system, and measures the quantum signal’s phase distortions, using the wavefront sensor of the adaptive optics system.
    \item The receiver compensates the quantum signal’s phase distortions, using the deformable mirror of the adaptive optics system, and enhances the quantum signal’s SNR, using the signal enhancer and recoverer module of the AQCA.
    \item The receiver measures the quantum signal, using a quantum detector, and applies the quantum code to the quantum signal, using the syndrome measurement and decoding steps of the QEC process, to recover the original quantum state.
    \item The transmitter and the receiver exchange classical information, such as basis choices, error correction codes, and privacy amplification keys, using a classical channel, such as a radio or optical link.
    \item The transmitter and the receiver perform post-processing steps, such as error correction, privacy amplification, and authentication, to ensure the security and reliability of the quantum communication.
\end{enumerate}


\section{LF QSDC Simulation Plan and Analysis}

In the forthcoming research phase, our primary objective is to thoroughly simulate and evaluate the performance of our quantum-aware LDPC codes, PAT technologies, and the atmospheric quantum correction algorithm. 

\subsection{Simulation Plan}

Figure 2 outlines the process of the simulation plan: 

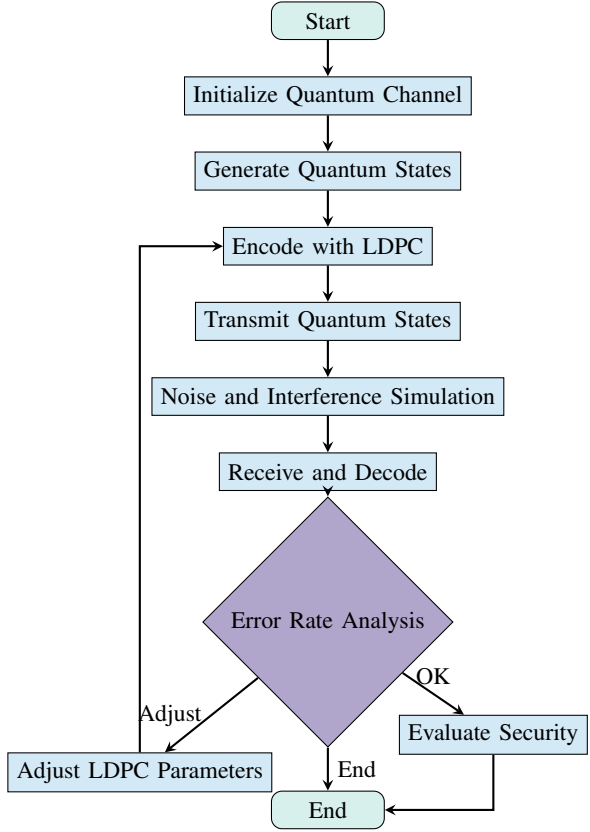
\begin{figure} [ht]
    \centering
    \begin{tikzpicture}[node distance=1cm, font = \small]
    \tikzset{
      startstop/.style={rectangle, rounded corners, minimum width=1.5cm, minimum height=0.5cm, text centered, draw=black, fill=color1!50, font=\small},
      io/.style={trapezium, trapezium left angle=70, trapezium right angle=110, minimum width=1.5cm, minimum height=0.5cm, text centered, draw=black, fill=color2!50, font=\small},
      process/.style={rectangle, minimum width=1.5cm, minimum height=0.5cm, text centered, draw=black, fill=color3!50, font=\small},
      decision/.style={diamond, minimum width=1.5cm, minimum height=0.5cm, text centered, draw=black, fill=color4!50, font=\small},
      arrow/.style={thick,->,>=stealth}
    }
    \node (start) [startstop] {Start};
    \node (init) [process,below of=start] {Initialize Quantum Channel};
    \node (gen) [process, below of=init] {Generate Quantum States};
    \node (encode) [process, below of=gen] {Encode with LDPC};
    \node (transmit) [process, below of=encode] {Transmit Quantum States};
    \node (noise) [process, below of=transmit] {Noise and Interference Simulation};
    \node (receive) [process, below of=noise] {Receive and Decode};
    \node (error) [decision, below of=receive, yshift=-1cm] {Error Rate Analysis};
    \node (adjust) [process, left of=error, xshift=-1.5cm, yshift=-2cm] {Adjust LDPC Parameters};
    \node (security) [process, right of=error,  xshift=1.2cm, yshift=-1.5cm] {Evaluate Security};
    \node (end) [startstop, below of=error, yshift=-1.5cm] {End};
    
    \draw [arrow] (start) -- (init);
    \draw [arrow] (init) -- (gen);
    \draw [arrow] (gen) -- (encode);
    \draw [arrow] (encode) -- (transmit);
    \draw [arrow] (transmit) -- (noise);
    \draw [arrow] (noise) -- (receive);
    \draw [arrow] (receive) -- (error);
    \draw [arrow] (error) -- node[anchor=east] {Adjust} (adjust);
    \draw [arrow] (adjust) |- (encode);
    \draw [arrow] (error) -- node[anchor=south] {OK} (security);
    \draw [arrow] (security) |- (end);
    \draw [arrow] (error) -- node[anchor=west] {End} (end);
    
    \end{tikzpicture}
    \caption{Flowchart outlining the overview of our simulation plan}
    \label{fig:simulation plan}
\end{figure}

The testing framework for the LF QSDC system meticulously evaluates its operational efficacy, commencing with the initialization of the quantum channel (\(\mathcal{Q}_C\)), a conduit essential for the transmission of entangled photons (\(\Psi_{\text{ent}}\)). This phase sets the groundwork for secure quantum communication by establishing a pathway for encoded quantum states (\(\Psi_{\text{enc}}\)) utilizing quantum-aware LDPC coding (\(C_{\text{LDPC}}\)). Such encoding is pivotal, aiming to bolster error correction capabilities without undermining the quantum states' coherence and security. 

The protocol then progresses to the transmission phase (\(\Phi_{\text{transmit}}\)), where encoded quantum states are dispatched through the free-space medium, encountering environmental noise (\(\mathcal{N}\)), simulating real-world atmospheric conditions. The subsequent reception and decoding phase is critical, employing \(C_{\text{LDPC}}\) to ameliorate errors introduced during transmission, highlighted by the quantum channel's intrinsic error characteristics (\(\epsilon_{\text{quantum}}\)). An in-depth error rate analysis (\(\eta_{\text{error}}\)) ensues, assessing the integrity of the received quantum information against the original transmission. This analysis is instrumental in driving the iterative optimization of LDPC parameters (\(\Theta_{\text{opt}}\)), with the goal of minimizing error rates and maximizing system fidelity (\(F_{\text{system}}\)).

Concluding the protocol, a rigorous security evaluation (\(\Sigma_{\text{security}}\)) is conducted to ensure the system's robustness against potential eavesdropping attempts, affirming the LF QSDC system's ability to preserve the secrecy and accuracy of the information sent.

\subsection{Predicted Results and Analysis}

This part aims to assess the theoretical efficacy of our suggested lossless free space and extended-range transmission methods, incorporating quantum-aware LDPC, PAT technologies, and the atmospheric quantum correction algorithm. 

Figure 3 shows the predicted secure information transmission rates of the proposed transmission scheme, the secure coding based on JEEC Coding \cite{b15}, the secure coding based on LPS codes, and Cs for a practical QSDC system, without the consideration of the loss caused by the delayed fiber \cite{b32}.

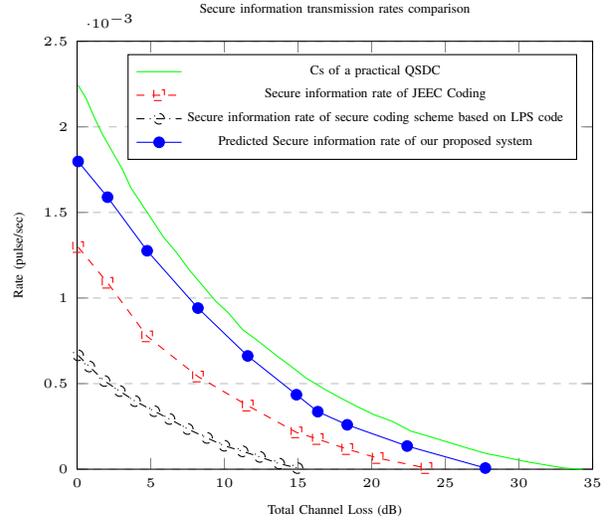
\begin{figure} [ht]
    \centering
    \begin{tikzpicture}
    \begin{axis}[
        title={Secure information transmission rates comparison},
        xlabel={Total Channel Loss (dB)},
        ylabel={Rate (pulse/sec)},
        xmin=0, xmax=35,
        ymin=0, ymax=2.5e-3,
        legend pos=north east,
        ymajorgrids=true,
        grid style=dashed,
        legend entries={
            Cs of a practical QSDC,
            Secure information rate of JEEC Coding,
            Secure information rate of secure coding scheme based on LPS code,
            Predicted Secure information rate of our proposed system
        },
        style={font=\tiny}
    ]
    
    \addplot[
        color=green,
        mark=none,
        solid
        ]
        coordinates {
        (0.06894841269841123, 0.0022434782608695652)
        (0.6205357142857135, 0.002165217391304348)
        (1.172123015873014, 0.0020543478260869564)
        (1.7237103174603163, 0.001956521739130435)
        (2.413194444444441, 0.0018521739130434782)
        (3.1026785714285694, 0.0017478260869565217)
        (3.65426587301587, 0.0016434782608695652)
        (4.4126984126984095, 0.0015456521739130435)
        (5.860615079365077, 0.0013565217391304349)
        (6.756944444444445, 0.0012652173913043478)
        (7.584325396825394, 0.001167391304347826)
        (8.549603174603172, 0.0010695652173913044)
        (9.376984126984125, 0.0009847826086956523)
        (10.273313492063492, 0.0009130434782608698)
        (11.238591269841267, 0.0008152173913043475)
        (12.272817460317457, 0.0007500000000000002)
        (13.513888888888888, 0.0006652173913043479)
        (14.34126984126984, 0.0006130434782608699)
        (15.513392857142852, 0.0005347826086956522)
        (16.754464285714278, 0.0004695652173913045)
        (17.857638888888886, 0.0004173913043478265)
        (19.029761904761905, 0.0003652173913043476)
        (20.132936507936506, 0.00031956521739130456)
        (21.37400793650793, 0.0002804347826086957)
        (22.68402777777777, 0.00022173913043478308)
        (23.994047619047613, 0.00018913043478260881)
        (25.235119047619044, 0.00015652173913043499)
        (26.476190476190474, 0.00012391304347826116)
        (27.786210317460316, 0.0000913043478260869)
        (28.95833333333333, 0.00007173913043478268)
        (30.406249999999993, 0.00004565217391304345)
        (31.716269841269835, 0.000026086956521738803)
        (33.095238095238095, 0.000006521739130434592)
        (34.336309523809526, 0)
        };
    
    \addplot[
        color=red,
        mark=square,
        dashed
        ]
        coordinates {
        (0.06894841269841123, 0.0012978260869565218)
        (2.0684523809523796, 0.0010891304347826086)
        (4.7574404761904745, 0.000776086956521739)
        (8.204861111111109, 0.0005413043478260868)
        (11.58333333333333, 0.0003717391304347826)
        (14.892857142857137, 0.00021521739130434805)
        (16.340773809523803, 0.0001760869565217392)
        (18.34027777777778, 0.00011739130434782613)
        (20.408730158730158, 0.00006521739130434766)
        (23.71825396825396, 0.000006521739130434592)
        };
    
    \addplot[
        color=black,
        mark=o,
        dashdotted
        ]
        coordinates {
        (0.06894841269841123, 0.0006652173913043479)
        (0.8273809523809508, 0.0005999999999999998)
        (1.8616071428571423, 0.000515217391304348)
        (2.895833333333332, 0.0004565217391304349)
        (3.930059523809522, 0.00039782608695652184)
        (5.240079365079362, 0.0003391304347826088)
        (6.274305555555552, 0.00029347826086956533)
        (7.515376984126982, 0.00023478260869565226)
        (8.825396825396824, 0.00018260869565217422)
        (9.99751984126984, 0.00013695652173913078)
        (11.238591269841267, 0.00010434782608695695)
        (12.410714285714283, 0.00007173913043478268)
        (13.720734126984125, 0.00003260869565217383)
        (14.961805555555552, 0.000006521739130434592)
        };
    
    \addplot[
        color=blue,
        mark=*,
        solid
        ]
        coordinates {
        (0.06894841269841123, 0.0017978260869565218)
        (2.0684523809523796, 0.0015891304347826086)
        (4.7574404761904745, 0.001276086956521739)
        (8.204861111111109, 0.0009413043478260868)
        (11.58333333333333, 0.0006617391304347826)
        (14.892857142857137, 0.00043521739130434805)
        (16.340773809523803, 0.0003360869565217392)
        (18.34027777777778, 0.00025939130434782613)
        (22.408730158730158, 0.00013521739130434766)
        (27.71825396825396, 0.000006821739130434592)
        };
    \end{axis}
\end{tikzpicture}
    \caption{Graph showcasing the predicted secure information transmission rates of the proposed transmission scheme, the secure coding based on JEEC Coding, the secure coding based on LPS codes, and Cs for a practical QSDC system, without the consideration of the loss caused by the delayed fiber}
    \label{fig: Simulation results}
\end{figure}

Our estimations for our proposed system's performance are substantiated by empirical evidence and mathematical proofs from recent research such as works by Hu et al. and Babar et al. and\cite{b30, b31}. Specifically, studies by Roffe \cite{b33} on quantum LDPC codes and by Ghilea \cite{b34} on quasi-cyclic multi-edge LDPC codes demonstrate significant advancements in error correction, decoder efficiency, and communication distance for quantum key distribution systems. Furthermore, Mele, Lami, and Giovannetti \cite{b35} Investigated how quantum communication technologies withstand noise and attenuation, laying a robust foundation for our forecasts. The results collectively highlight the capability of our suggested system to improve upcoming QSDC systems, providing robust empirical and mathematical backing for our calculations.

The forecasts indicate that our system will improve the operational benchmark for QSDC systems by showcasing its capability to maintain elevated secure information speeds, even amidst considerable channel loss. Yet, the present version of our system remains incapable of facilitating effective long-range QSDC communications. However, our upcoming improvements and updates to the coding of our system will concentrate on enhancing error rectification and adjusting to the unique difficulties of quantum channels, like quantum noise and decoherence. The enhancements are set to align our system with the rigorous criteria of practical QSDC.


\section{Implementation Plan for LF QSDC}

\subsection{Technical Implementation of LF QSDC}

\subsubsection{1-Way Transmission Protocols in Free-Space Channel}

For extensive free-space communication, the one-way transmission methods (RECON protocol and QKPC protocol) might offer more benefits than the two-way protocols, given the various elements in the free-space channel influencing communication \cite{b36}. The conveyance of data across extensive distances via a free-space channel is fraught with multiple difficulties. Factors such as beam-spreading, atmospheric turbulence, absorption and scattering, background light (sunlight), geometrical loss,
and weather conditions could all affect communication \cite{b37,b38,b39}. Figure 4 summarizes the main characteristics of the free-space channel and their impacts on optical signal transmission. 

\begin{figure} [ht]
    \centering
    \includegraphics[scale = 0.5]{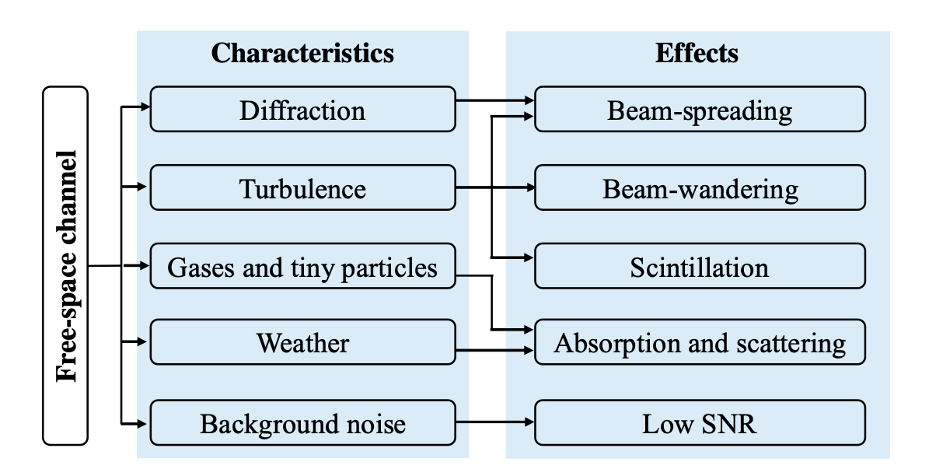}
    \caption{Summarizes the main characteristics of free-space channel and their impacts on optical signal transmission\cite{b62}}
    \label{fig:enter-label}
\end{figure}

Consequently, one-way transmission methods like RE-CON or QKPC protocols might offer greater benefits than the two-way approaches used in LF QSDC \cite{b40}. Protocols for one-way transmission simplify the process by removing the necessity for the signal to return to the sender for additional processing, thus decreasing the travel distance of quantum states \cite{b41,b42}. 

\subsubsection{Experimental Implementation}

In 2020, there was a reported successful experimental setup of the free-space DL04 QSDC protocol. The setup, as shown in Figure 5, achieved a data transmission rate of 500 bits per second across a 10-meter free-space channel with a notably low Quantum Bit Error Rate (QBER) of $0.49\% \pm 0.27\%$\cite{b62}.

Attaining a low QBER in experiments is a key measure of QSDC’s resilience and dependability in free-space channels \cite{b44}. A minimal QBER indicates the ability to convey quantum data with great accuracy, crucial for ensuring secure communication demonstrated by both works in QSDC and QKD \cite{b45,b46,b47}. Attaining a data transfer speed of 500 bits per second across a 10-meter free-space channel showcases the protocol’s capability for effective data transfer \cite{b62}. 

The experimental arrangement employed a phase encoding technique to transmit quantum states in a vacuum. In this method, Bob introduces one of four possible phases {0, $\pi$/2, $\pi$, 3$\pi$/2}  to each pulse passing through a longer optical path, creating four distinct phase-encoded quantum states. Employing a phase-encoding approach in the experiment, enabling the generation of four unique phase-encoded quantum states, demonstrates the real-world use of advanced quantum communication methods in open-space settings \cite{b64}. 

\begin{figure*} [ht]
    \centering
    \includegraphics[scale=0.4]{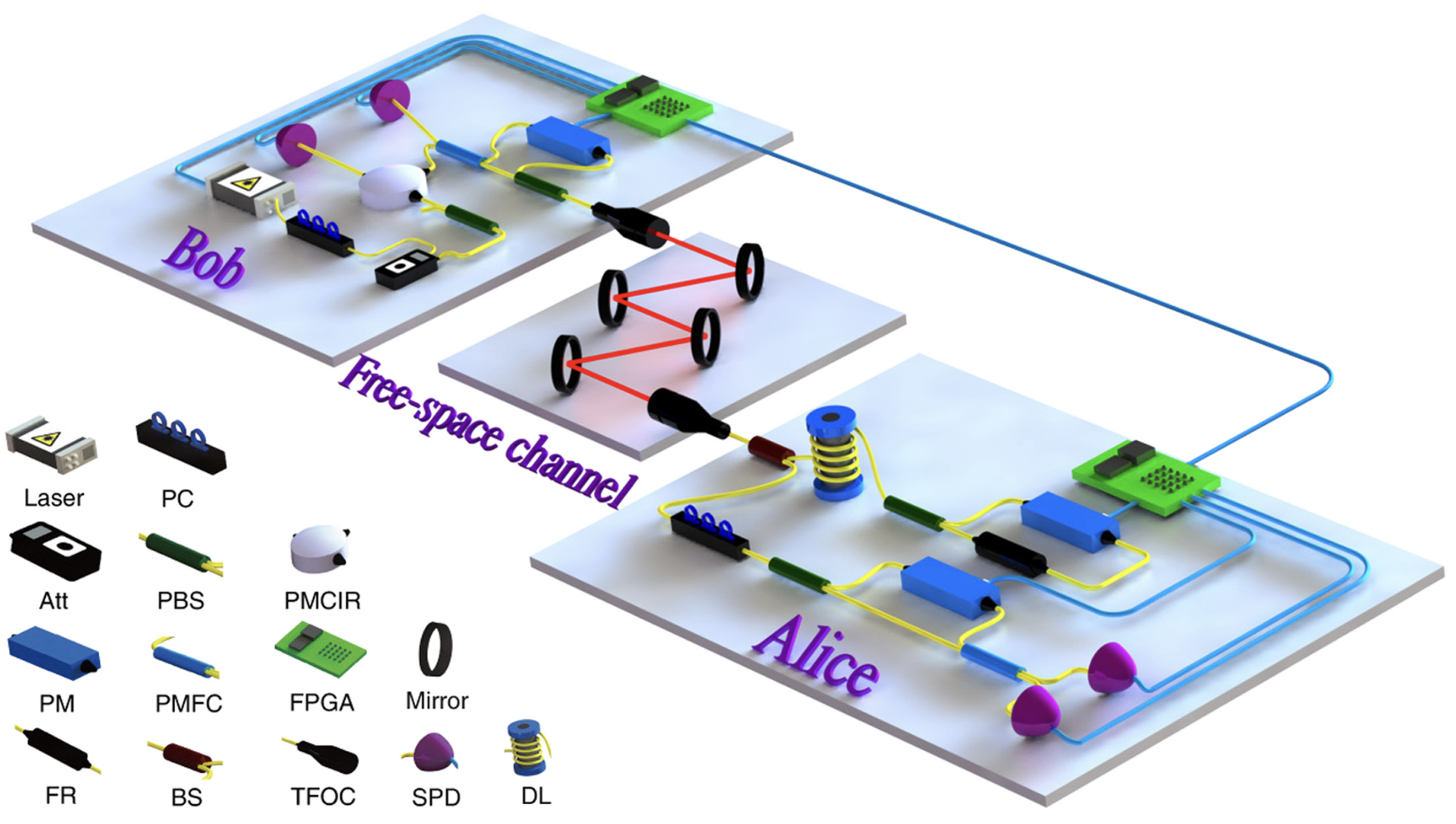}
    \caption{Summary of the main characteristics of the free-space channel and their impacts on optical signal transmission\cite{b62}}
    \label{fig:experimental-setup}
\end{figure*}
\color{black}
Included in the experimental arrangement were systems for identifying eavesdropping, an essential element of secure quantum communication. The protocol's capacity to adjust pulses for security verifications and secure information encoding showcases its effectiveness in maintaining communication confidentiality and integrity, even amidst possible risks\cite{b62}. For LF QSDC, this characteristic is crucial, given the paramount importance of safeguarding communication from eavesdropping.

Included in the experimental arrangement were also systems for identifying eavesdropping, an essential element of secure quantum communication. A protocol’s capacity to adjust pulses for security verifications and secure information encoding showcases its effectiveness in maintaining communication confidentiality and integrity is crucial and achieveavle, even amidst possible risks as demonstrated by this experiment and Pan et al.\cite{b43}. 

The success of these experiments confirms the practicality of LF QSDC, showcasing the protocol’s capacity to convey quantum data with great accuracy and safety via free-space channels.

\subsection{Feasibility and Adaptability of Satellite-based QSDC}

\subsubsection{Feasibility of Satellite Communication with LF QSDC}

Earlier studies on the satellite-to-ground transfer of quantum states have yielded encouraging outcomes, with QBERs remaining within the tolerable limits for safe quantum communication\cite{b48,b49,b50}. The ability to transmit effectively across distances from 2,200 km to more than 36,000 km, maintaining link losses in the range of 100 to 110 dB, demonstrates that basic physics enables quantum communication across the extensive distances needed for satellite communication\cite{b50}. 

The DL04 protocol, devoid of memory, plays a crucial role in LF QSDC by simplifying the processes of quantum state storage and management \cite{b19}. The DL04 protocol’s LF flexibility in adapting to free-space environments renders it an appealing choice for satellite communication, especially in addressing the difficulties of transmitting quantum signals across Earth’s atmosphere. The incorporation of quantum-aware LDPC coding into LF QSDC markedly enhances the system’s error rectification abilities, tackling the problems caused by extensive transmission and atmospheric influences \cite{b51,b52,b53}. Furthermore, the use of PAT technologies helps sustain a consistent and precise communication connection between 
satellites and terrestrial stations \cite{b54}.

\subsubsection{Expanding Feasibility with LF QSDC}

Enhancing the feasibility of LF QSDC involves comprehensive advancements across various technical and regulatory dimensions. Key initiatives include refining beam propagation and encoding methods to counter atmospheric disturbances and environmental disruptions as shown in Figure 5, ensuring high communication accuracy and low Quantum Bit Error Rate (QBER) across 
vast distances \cite{b55,b56}. Additionally, integrating quantum repeaters and relay satellites expands the scope of QSDC, enabling a global quantum communication network that overcomes spatial limitations inherent in direct quantum communication \cite{b57}. Improvements in daylight operation capabilities through advanced detectors and narrowband filters are crucial for QSDC's effectiveness in satellite applications, addressing challenges posed by solar radiation \cite{b58}.

Further, merging QSDC systems with existing satellite communication infrastructures requires addressing payload capacity, energy needs, and upgrade compatibility, ensuring effective implementation of quantum communication methods \cite{b59}. Promoting standardization and interoperability among quantum communication systems through uniform protocols and error-correcting techniques is essential for establishing a scalable and unified global network \cite{b60}. Developing regulatory and policy frameworks is also critical for overseeing the application and ensuring the security and privacy of satellite-based quantum communication, addressing potential legal and security issues \cite{b61}. These collective efforts will significantly enhance the practicality of satellite-based LF QSDC, paving the way for a secure and efficient global quantum communication network.

\subsection{Web 3.0 Compatibility}

The integration of LF QSDC with Web 3.0 technologies marks a pivotal advancement in securing decentralized networks through quantum communication methods. This collaboration aims to create a secure transactional layer within Web 3.0, enabling nodes to directly and securely exchange data, thereby enhancing transaction secrecy and reliability against both conventional and quantum cryptographic threats \cite{b62,b63,b64}. Furthermore, the adaptation of LF QSDC within Web 3.0 involves developing quantum-resistant smart contracts, ensuring their execution and outcomes are secured through quantum communication, thus requiring smart contract systems to support encryption and verification processes compatible with LF QSDC protocols \cite{b65,b66,b67}.

Additionally, LF QSDC contributes to Web 3.0's goals of user privacy and anonymity by establishing secure, direct communication channels that eliminate the need for intermediaries in key exchanges or transaction verifications. This approach not only strengthens confidentiality within the network but also introduces an additional layer of transaction security based on quantum mechanics, offering a robust defense against potential quantum cryptographic attacks \cite{b68}.

For LF QSDC's successful integration into Web 3.0, it must align with existing protocols, modifying transaction verification systems to accommodate quantum-secured information and ensuring the ledger supports and accurately documents quantum-secured transactions \cite{b69,b70,b71}. Additionally, considering the increasing transaction volumes on Web 3.0 networks,

\subsection{Stages of Gradual Implementation}
Figure 6 summarizes the overview of the stages of gradual implementation:

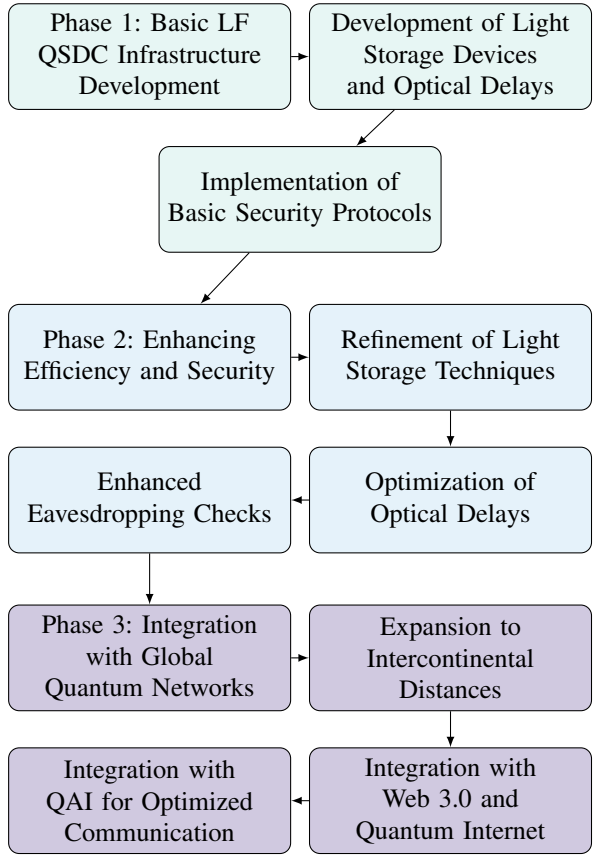
\begin{figure} [ht]
    \centering

\begin{tikzpicture}[
    node distance=1.5cm,
    auto,
    block/.style={
        rectangle,
        draw,
        text width=10em,
        text centered,
        rounded corners,
        minimum height=4em,
        fill=color1!30
    },
    block1/.style={
        rectangle,
        draw,
        text width=10em,
        text centered,
        rounded corners,
        minimum height=4em,
        fill=color3!30
    },
    block2/.style={
        rectangle,
        draw,
        text width=10em,
        text centered,
        rounded corners,
        minimum height=4em,
        fill=color4!30
    },
    line/.style={
        draw, 
        -latex
    }
]
\tikzstyle{startstop} = [rectangle, rounded corners, minimum width=1.5cm, minimum height=0.5cm, text centered, draw=black, fill=red!30, font=\small]
\tikzstyle{io} = [trapezium, trapezium left angle=70, trapezium right angle=110, minimum width=1.5cm, minimum height=0.5cm, text centered, draw=black, fill=blue!30, font=\small]
\tikzstyle{process} = [rectangle, minimum width=1.5cm, minimum height=0.5cm, text centered, draw=black, fill=orange!30, font=\small]
\tikzstyle{decision} = [diamond, minimum width=1.5cm, minimum height=0.5cm, text centered, draw=black, fill=green!30, font=\small]
\tikzstyle{arrow} = [thick,->,>=stealth]

\node [block] (init) {Phase 1: Basic LF QSDC Infrastructure Development};
\node [block, right of=init, xshift= 2.5cm] (technologies) {Development of Light Storage Devices and Optical Delays};
\node [block, below of=technologies, xshift= - 2cm, yshift=-0.4cm] (securityProtocols) {Implementation of Basic Security Protocols};

\node [block1, below of=securityProtocols, xshift= -2cm, yshift =-0.6cm] (phase2) {Phase 2: Enhancing Efficiency and Security};
\node [block1, right of=phase2,xshift= 2.5cm] (storageOptimization) {Refinement of Light Storage Techniques};
\node [block1, below of=storageOptimization, yshift=-0.4cm] (delayOptimization) {Optimization of Optical Delays};
\node [block1, left of=delayOptimization,xshift= - 2.5cm] (eavesdropEnhancements) {Enhanced Eavesdropping Checks};

\node [block2, below of=eavesdropEnhancements, yshift=-0.6cm] (phase3) {Phase 3: Integration with Global Quantum Networks};
\node [block2, right of=phase3,xshift= 2.5cm] (networkExpansion) {Expansion to Intercontinental Distances};
\node [block2, below of=networkExpansion, yshift=-0.4cm] (webIntegration) {Integration with Web 3.0 and Quantum Internet};
\node [block2, left of=webIntegration, xshift= - 2.5cm] (qaiIntegration) {Integration with QAI for Optimized Communication};

\path [line] (init) -- (technologies);
\path [line] (technologies) -- (securityProtocols);
\path [line] (securityProtocols) -- (phase2);
\path [line] (phase2) -- (storageOptimization);
\path [line] (storageOptimization) -- (delayOptimization);
\path [line] (delayOptimization) -- (eavesdropEnhancements);
\path [line] (eavesdropEnhancements) -- (phase3);
\path [line] (phase3) -- (networkExpansion);
\path [line] (networkExpansion) -- (webIntegration);
\path [line] (webIntegration) -- (qaiIntegration);
\end{tikzpicture} \newline
    \caption{Flowchart showing the overview of the stages of gradual implementation}
    \label{fig: Stages if Gradual Implementation}
\end{figure}

\subsubsection{Phase 1: Development of Basic LF QSDC Infrastructure}

The initial configuration focuses on establishing a basic system to create, store, and transmit entangled photon pairs between two locations, Alice and Bob, requiring optical systems like photon sources, detectors, and communication channels. Future efforts will explore light storage devices using electromagnetic transparency and optical delays for sequential EPR pair production, aiming to integrate these technologies into LF QSDC and set up security measures to detect eavesdropping and evaluate secure communication efficiency.

\subsubsection{Phase 2: Enhancing Efficiency and Security}

With the advancement of technology, enhance and fine-tune methods for light storage to prolong photon retention while maintaining their quantum conditions. The advancement is vital for broadening the real-world application of LF QSDC. Enhance the application of optical lags to guarantee accurate synchronization in the transmission of the C-sequence and M-sequence. This includes fine-tuning the delay \( \tau \) to accommodate different distances and transmission rates, ensuring that the system can operate efficiently over long distances. Enhance the protocols for eavesdropping checks by incorporating more sophisticated quantum measurements and real-time analysis. This improvement aims to reduce error rates and increase the system's resilience against potential quantum attacks.

\subsubsection{Phase 3: Integration with Global Quantum Networks}

This stage entails partnering with current quantum communication endeavors to establish a cohesive and secure network of quantum communication. Combine LF QSDC with burgeoning Web 3.0 technologies and the Quantum Internet, facilitating protected, immediate quantum communication for diverse uses, ranging from secure messaging to quantum cloud computing.

\subsection{Business Case of LF QSDC in The Web 3.0 Era}

Integrating LF QSDC with Web 3.0 represents a significant advancement in securing decentralized networks against quantum computing threats, emphasizing its transformative potential for Web 3.0's security and trust mechanisms in an increasingly digital and quantum-aware world [56]. This integration addresses the vulnerabilities of traditional cryptographic methods to quantum computing by offering a quantum mechanics-based solution for secure, direct communication over long distances without intermediaries, significantly enhancing Web 3.0's security posture and ensuring sustained trust and reliability \cite{b62}.

The necessity for quantum-resistant security protocols in Web 3.0 is increasingly critical, with LF QSDC fulfilling this requirement by providing a secure, scalable, and efficient framework. This advancement not only protects against future quantum threats but also reinforces defenses against sophisticated conventional attacks, positioning Web 3.0 networks at the cutting edge of secure digital technology. The integration of LF QSDC into Web 3.0 is especially pertinent as the adoption of Web 3.0 expands across various sectors, highlighting the growing demand for robust security solutions in the face of quantum computing's rise and offering a competitive advantage to early adopters \cite{b62}.

However, integrating LF QSDC into Web 3.0 comes with challenges, including technological complexity and interoperability issues. Strategic partnerships, ongoing research and development, and adaptable integration strategies are essential for mitigating these risks and staying abreast of quantum computing advancements. Despite initial costs related to research, technology acquisition, and system upgrades, the long-term benefits of enhanced security and network durability justify the investment, promising increased user confidence and new opportunities in quantum-secure technologies \cite{b62}. The strategic incorporation of LF QSDC into Web 3.0 not only fortifies network security but also positions Web 3.0 networks as leaders in the next era of digital innovation, ready for the challenges of the quantum computing age.


\section{Discussion}

\subsection{Comparison With Similar Protocols}

\subsubsection{Security}

LF QSDC leverages quantum mechanics' inherent principles to offer a groundbreaking level of security, making it particularly resilient against both eavesdropping and sophisticated quantum attacks. This protocol benefits from quantum phenomena such as the no-cloning theorem and entanglement, ensuring that any interception attempt would alter the quantum state, alerting the communicating parties to a potential security breach. Unlike Quantum Key Distribution, LF QSDC transmits encrypted data directly without the need for key exchanges, providing enhanced security against quantum and classical threats and making it highly suitable for secure communications in the decentralized nature of the Web 3.0 era. LF QSDC and DL04 both offer high levels of security based on quantum principles; however, LF QSDC's emphasis on long-distance free-space communication adds an extra layer of complexity and potential vulnerability due to atmospheric effects, which it addresses with advanced PAT systems \cite{b72,b73}. Additionally, LF QSDC does not require pre-shared secret keys, offering a theoretical security advantage over other protocols such as RECON protocol, which depends on the secure distribution of initial keys.

\subsubsection{Distance}

Yin explored free-space QSDC's capability for long-distance communication through free-space channels and successfully demonstrated quantum teleportation over a 97-km channel and entanglement distribution over a 101.8-km two-link channel, achieving an average fidelity of 80.4\% for six initial states despite significant channel loss (35-53 dB for teleportation and 66-85 dB for entanglement distribution)\cite{b74}. This showcases LF QSDC's potential for satellite-based quantum communication, significantly extending the feasible communication distance beyond traditional QSDC methods. Achieving free-space QSDC over distances of 100 kilometers has been validated through other successful experiments in quantum teleportation and the distribution of entanglement across distances surpassing 100 km\cite{b8,b75}.

\subsubsection{Channel Loss and Atmospheric Disturbances}

LF QSDC faces significant challenges related to channel loss and atmospheric disturbances, particularly over long distances. Yin highlights the critical role of the PAT systems in overcoming these challenges, such as atmospheric turbulence, which can severely affect the stability and accuracy of quantum communication. The PAT system's ability to maintain high tracking accuracy, better than 3.5 µrad over a 97 km free-space link, is essential for ensuring that the quantum signal remains aligned with the receiver despite atmospheric turbulence and other disturbances. Additionally, the inclusion of coarse and fine tracking systems, controlled by a close-loop via the telescope's own rack and piezo ceramics at the receiver's side, aims to reduce low-frequency shaking caused by ground settlement and passing vehicles, achieving an average fidelity of 80.4\% over a 35-53 dB loss quantum channel\cite{b74}.

\subsubsection{Cost}

Implementing LF QSDC, especially for applications such as global Web 3.0 networks, involves sophisticated technology and infrastructure. This includes satellites, ground stations, and advanced tracking technologies, making it more expensive compared to other QSDC protocols that operate over shorter distances or through fiber channels. The investment in technology and infrastructure signifies a considerable cost, which is a critical consideration for the widespread adoption of LF QSDC in various applications.

\subsection{Limitations and Future Work}

The advent of Free-Space Long-Distance QSDC is poised to revolutionize the QSDC landscape by facilitating secure quantum communication across vast distances, unencumbered by the physical limitations of fiber-optic systems. This innovation is expected to have a profound impact on the implementation and scalability of QSDC technologies, enabling a seamless and secure global quantum communication network.

LF QSDC's integration into Web 3.0 infrastructure signifies a monumental shift towards creating a decentralized and secure internet, where the intrinsic security features of quantum communication can protect against ever-evolving cyber threats. The ability of LF QSDC to operate in free space allows for the establishment of quantum links between any two points on the globe, directly supporting the foundational principles of Web 3.0 by enhancing data integrity, security, and privacy across decentralized networks.

Furthermore, LF QSDC's potential to extend the reach of quantum networks without the need for extensive physical infrastructure paves the way for more inclusive access to quantum communication technologies. This democratization of technology is crucial for leveraging the full potential of quantum advancements in various sectors, including secure communications, distributed computing, and beyond, marking a significant leap toward a quantum-integrated future.

However, it is important to note that our model is a preliminary theoretical proposal of an LF QSDC system which comes with its own limitations and challenges as discussed below:

\subsubsection{Theoretical Challenges}

\begin{itemize}
    \item Quantum Decoherence and Noise: Quantum states are highly susceptible to decoherence and environmental noise, which can severely limit the distance over which LF QSDC can be effectively implemented.
    \item Security Proofs: Complete, universally accepted security proofs for LF QSDC under all potential attack scenarios are challenging to develop, raising concerns about its absolute security.
\end{itemize}

\subsubsection{Technological Constraints}

\begin{itemize}
    \item Quantum Sources and Detectors: The efficiency and reliability of quantum sources (e.g., single-photon sources) and detectors are critical, yet current technologies may not provide the necessary performance for long-distance communication.
    \item Atmospheric Interference: Free-space transmission is significantly affected by atmospheric conditions (e.g., cloud cover, atmospheric turbulence), which can degrade the quantum signal over long distances.
\end{itemize}

\subsubsection{Implementation Challenges}

\begin{itemize}
    \item Infrastructure Development: Establishing a global LF QSDC network requires significant investment in both ground-based and potentially satellite-based infrastructure, posing a substantial financial and logistical challenge.
    \item Interoperability: Compatibility with existing communication technologies and standards is crucial for widespread adoption, necessitating complex integration efforts.
\end{itemize}

\subsubsection{Impact Considerations}

\begin{itemize}
    \item Scalability: While promising for point-to-point communication, scaling LF QSDC to a multi-node quantum network presents considerable technical challenges.
    \item Accessibility: The high cost and complexity of LF QSDC technology may limit its accessibility, particularly in developing regions, potentially exacerbating the digital divide.
    \item Regulatory and Ethical Issues: The deployment of LF QSDC might raise questions regarding regulation, data sovereignty, and privacy, requiring careful consideration and potentially new legal frameworks.
\end{itemize}


\section{Conclusion}

This document delves into the integration of distant, free-space quantum secure direct communication in the digital age of Web 3.0, introducing a theoretical model aimed at enhancing the security aspects of the worldwide networking structure in the quantum age. This document highlights LF QSDC's capability to strengthen the Web 3.0 framework against various cryptographic dangers, including quantum and classical, by utilizing direct quantum communication, bypassing traditional key exchange methods. 

LF QSDC's core is based on the memory-free DL04 protocol, implemented by merging quantum-aware, low-density parity-check codes, sophisticated pointing, acquisition, and tracking technologies, along with algorithms for atmospheric quantum correction. These factors play a crucial role in overcoming current environmental and technical challenges that hinder the effective use of quantum communication technologies, especially in the realm of long-distance free-space communication. Additionally, we suggest a strategic plan that includes the development of quantum communication technologies, integrating them with the current Web 3.0 framework, and overcoming environmental and technical hurdles to ensure a secure and efficient data transmission channel. 

Implementing LF QSDC in Web 3.0 networks presents intricate engineering hurdles and necessitates significant progress in quantum communication technology. Upcoming studies might focus on creating advanced quantum error correction methods to improve the accuracy of quantum data transmission. Furthermore, investigating cutting-edge PAT systems to enhance the stability and precision of quantum signal alignment might greatly aid in the practicality of LF QSDC. Developing scalable quantum network structures for effortless integration with current Web 3.0 systems is another vital field for future research, guaranteeing that quantum security improvements do not hinder the network's operational efficiency or accessibility.


\section{Acknowledgement}

\textbf{Author Contributions:} These authors contributed equally: Yew Kee Wong, Yifan Zhou, Xinlin Zhou, Yan Shing Liang, and Zi Yan Li.

Under the guidance of Yew Kee Wong, Yifan Zhou, Yan Shing Liang, Xinlin Zhou, and Zi Yan Li conducted the research with Yifan Zhou and Yan Shing Liang leading the methodology development. The original draft was primarily written by Yifan Zhou, with significant contributions from Xinlin Zhou, Zi Yan Li, and Yan Shing Liang. Xinlin Zhou also undertook the task of reviewing and editing the manuscript. Yew Kee Wong supervised the project, provided revisions, and managed project administration.


\section{Data Availability}

The datasets analyzed during the current study are available from the corresponding author on reasonable request.

\end{document}